# Boosting Extra-functional Code Reusability in Cyber-physical Production Systems: The Error Handling Case Study

Birgit Vogel-Heuser, Juliane Fischer, Dieter Hess, Eva-Maria Neumann, Marcus Würr

**Abstract**— Cyber-Physical Production Systems (CPPS) are long-living and mechatronic systems, which include mechanics, electrics/electronics and software. The interdisciplinary nature combined with challenges and trends in the context of Industry 4.0 such as a high degree of customization, small lot sizes and evolution cause a high amount of variability. Mastering the variability of functional control software, e.g., different control variants of an actuator type, is itself a challenge in developing and reusing CPPS software. This task becomes even more complex when considering extra-functional software such as operating modes, diagnosis and error handling. These software parts have high interdependencies with functional software, often involving the human-machine interface (HMI) to enable the intervention of operators. This paper illustrates the challenges in documenting the dependencies of these software parts including their variability using family models. A procedural and an object-oriented concept for implementing error handling, which represents an extra-functional task with high dependencies to functional software and the HMI, are proposed. The suitability of both concepts to increase the software's reusability and, thus, its flexibility in the context of Industry 4.0 is discussed. Their comparison confirms the high potential of the object-oriented extension of IEC 61131-3 to handle planned reuse of extra-functional CPPS software successfully.

## 1 Introduction and Motivation

Cyber-Physical Production Systems (CPPS) are mechatronic, variant-rich and long-living systems connected to digital networks to use globally available data and services. Their development involves different disciplines, including mechanics, electrics/electronics and software engineering. An increasing amount of their functionality is realized by control software [1], which highly depends on the used automation hardware and the CPPS layout. Thus, software variants arise due to variable automation hardware, shortened production cycles and continuously changing customer requirements [2], stressing the need for systematic software reuse to enable flexible production in the context of Industry 4.0. Thereby, achieving a high degree of modularization via functionality encapsulation is an established strategy to derive high-quality, reusable software parts with standardized module interfaces that can be flexibly combined for controlling CPPS [3]. For functional software, e.g. the control of actuators, modular software structures are successfully applied. However, extra-functional software implementing communication tasks, diagnosis, error handling and operating modes is an essential part of control software: extra-functional code makes up around 50-75% of industrial control code [4] causing complexity in the control software [5] and thus reducing maintainability and consequently impacting long-term cost and the ability to innovate. The close dependencies between functional and extra-functional code make reuse and variant management of these software parts even more difficult. To illustrate the link of the human-machine interface (HMI) to (extra-)functional control software implemented in accordance to IEC 61131-3 on a programmable logic controller (PLC), the extra-functional task error handling is chosen. Error handling requires close connections to functional software parts, e.g., to trigger the emergency stop of an actuator in case of an error, and might also reference other extra-functional tasks such as operating mode change. Further, errors are communicated from the PLC to the HMI to enable operators to intervene and resolve an error if needed. This interaction of a human operator with a CPPS in case of an occurring error is explained from an application perspective in Fig. 1. Due to an error situation, the operator pushes the emergency stop of the CPPS, e.g., a machine being part of a CPPS (cf. Fig. 1, step 1). In the PLC, the operating mode is changed accordingly, and the machine is stopped either immediately or in controlled steps (2). The HMI control panel displays the machine status and an error message to the operator (4), which the PLC previously transmitted to the HMI control panel (3). Depending on the operator's reaction, e.g., removing a jammed work piece manually (5), subsequently

- *Birgit Vogel-Heuser, Juliane Fischer and Eva-Maria Neumann are with the Institute of Automation and Information Systems, Technical University of Munich, Boltzmannstr. 15, Garching near Munich, Germany. Birgit Vogel-Heuser is also Core Member of MDSI and Member of MIRMI. E-mail: {vogel-heuser |juliane.fischer | eva-maria.neumann}@tum.de.*
- *Dieter Hess is with the CODESYS GmbH, Memminger Str. 151, 87439 Kempten (Allgäu), Germany. E-mail: d.hess@codesys.com.*
- *Marcus Würr is with the Schneider Electric Automation GmbH, Schneiderplatz 1, 97828 Marktheidenfeld, Germany. E-mail: marcus.wuerr@se.com.*

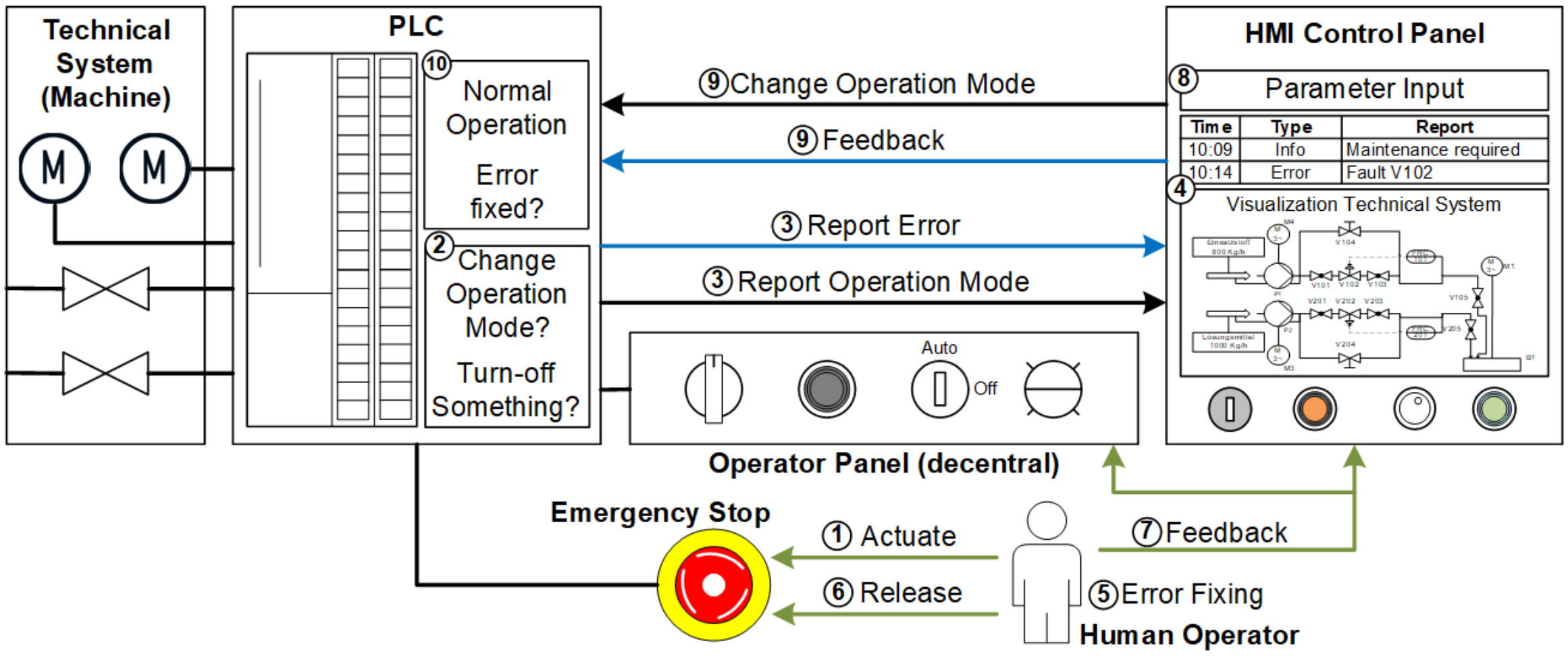

Fig. 1. Relationship between opertors interacting via HMI and PLC in case of an error (emergency stop actuation by operator, CPPS reaction until return to automatic mode) in accordance to [6]

releasing the actuated emergency stop switch (6) and then acknowledging the error situation via HMI control panel (7, right fork, 8, 9) or a decentralized operator panel (7, left fork), the PLC checks its status (10). If necessary, the PLC allows an organized restart in manual mode and a subsequent manual change to automatic mode if all interlocks are correctly parameterized. Alternatively, the operating mode can also be set back to automatic mode after the operator has acknowledged the error on the HMI panel (8), which is checked by the PLC after transmission (9, 10).

This simple example shows that error handling, operating modes, and HMI are strongly interlinked, but they have fundamentally different implementation and code structures. This makes it difficult to manage their variability and systematic reuse. In computer science object-oriented (OO) programming is successfully applied to enable encapsulation, standardized interfaces and reuse. Although the last update of the PLC programming standard IEC 61131-3 enables OO for PLCs, it is not yet established in industry and guidelines for its application are rare [7]. To bridge this gap and illustrate the challenges and potential solutions to deal with extra-functional software parts, this paper gives an introduction to the design decisions required to implement *error handling*. Further, two concepts – a procedural and an OO one – for its reusable implementation are presented and compared, which highlights the great potential of OO to manage extra-functional software.

The remainder of this paper is structured as follows: Sec. 2 provides an overview of the state of the art regarding reuse and variant management of control software, including background information on extra-functional control software and OO. Next, Sec. 3 presents software design decisions related to error handling and points out means of documenting variability as a basis for planned software reuse utilizing a lab-sized demonstrator. Subsequently, Sec. 4 introduces a procedural and an OO concept for error handling and closes with a comparison of the two presented concepts. Finally, a short summary and outlook are provided in Sec. 5.

## 2 State of the Art – Reuse and Variants of (Extra-)Functional Control Software

First, challenges regarding the reuse of variant-rich, functional control software, including approaches for variant management, are presented in Sec. 2.1. Next, extra-functional control tasks in CPPS, which have high interdependencies with functional software and pose an additional challenge to software reuse, are introduced. Finally, Sec. 2.3 provides an introduction to the OO concepts of IEC 61131-3 as a new paradigm to ease software reuse and gives an overview of its current application in industry.

### 2.1 Challenges in Reusing Variant-rich, Functional Control Software from CPPS

Most CPPS are controlled by real-time capable PLCs programmed with textual and graphical languages defined in IEC 61131-3. A PLC software project contains global variables and Program Organization Units (POUs) encapsulating functionality as reusable software modules. Each POU consists of a declaration and an implementation part. Its functionality can be enclosed even more fine-grained within Actions that represent sub-functionalities of a POU. Three POU types, i.e., Programs, Function Blocks (FBs) and Functions, are distinguished, with FBs containing a simple class concept and maintaining an internal state [8]. During software design, general quality requirements, e.g., reliability, performance efficiency, compatibility, portability and maintainability should be considered [9]. Further, control software highly depends on the used automation hardware including its variability and on the desired functionality or customer-specific process logic. For example, a sorting conveyor with separators to sort work pieces can implement different sorting algorithms with the same automation hardware. Concluding, the variability of CPPS needs to be considered from different views, e.g., customer, mechanical and software view, which are linked and contain partially overlapping information.

In the software engineering domain, Software Product Line Engineering (SPLE) is a well-established approach for planned reuse of variant-rich software systems, where artifacts common to all variants are ideally implemented only once [10]. For representing an SPL as 150%-model including all implementation artifacts, Family Models can be used. They depict the software artifacts hierarchically and grouped into variability categories, i.e., mandatory (present in all variants), alternative (can be used interchangeably) and optional (present in only some variants). Figure 2 shows an example of two family models, which document the variability of a mono- and a bistable pneumatic cylinder typically used in assembly machines from hard- and software perspective. While both cylinders have the action *ACT_Extend* to extend (mandatory) via Valve 1 and the corresponding variable *DO_Extend*, the monostable cylinder retracts mechanically with a spring in contrast to the bistable cylinder, which requires an additional action *ACT_Retract* to control its second valve (optional). Thereby, the interdependencies between hard- and software pose additional challenges on variant management, since the different views and stakeholders, including their perception of the system's variability, have to be considered. Further, the variability in the control hardware influences the variability in the software regarding functional hardware control.

Available SPLE approaches for PLC software apply reverse engineering to document the variability and enable planned reuse of legacy software, e.g., Hinterreiter et al. [11], Schlie et al. with a metric-based approach [12], or ECCO aiming at the support of an enhanced application of *copy, paste and modify* for development and maintenance [13]. However, in CPPS the hardware highly influences the software's variability. Thus, documenting software variants only is not sufficient, as the variability's cause is missing. SPLE approaches, which target the interdisciplinary character of CPPS, are in theory available, such as [14-16]. However, they are not yet applicable in the industry due to challenges introduced by varying control logic and the dependencies to extra-functional aspects such as error-handling, which is additionally highly dependent on factors like the customer premises.

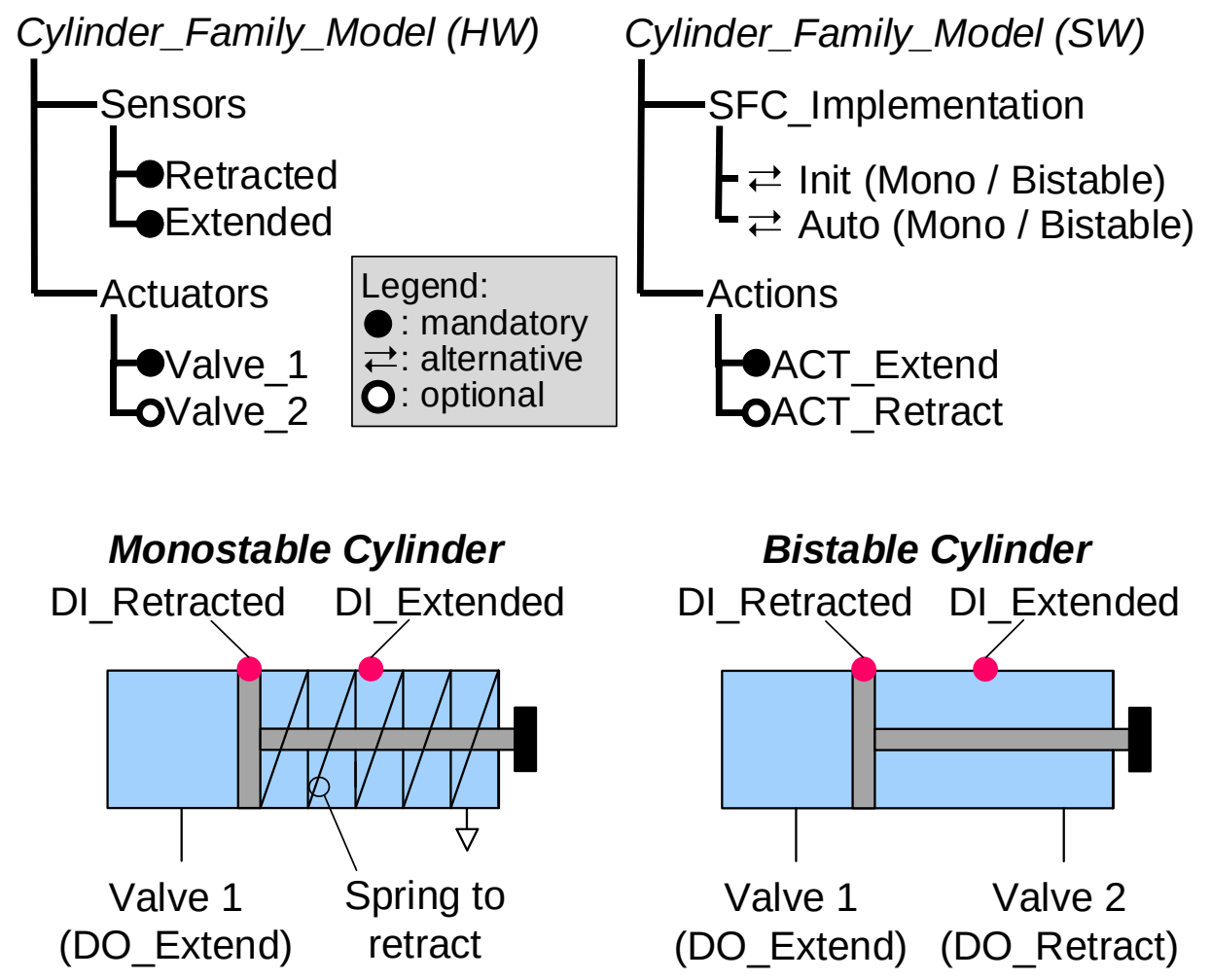

Fig. 2. Excerpt of Hardware and Software Family Models depicting the variability of mono- and bistable cylinders, adpoted from [6]

## 2.2 Reuse of extra-functional CPPS Software

Functional software for implementing the CPPS behavior in automatic mode is only a small part of the control logic compared to the extra-functional software, which is with 50-75% the much larger part [4]. Accordingly, Güttel et al. highlight that well-defined control software suitable for reuse, e.g., a POU for hardware control, includes standardized POU interfaces to HMI, various operating modes and diagnostic options [17]. Further, error detection, diagnosis and operating modes are implemented in about 50% of all software modules intended for planned reuse [18]. Thus, the following subsections present reuse approaches targeting different extra-functional aspects of CPPS software, which are also often referred to as software infrastructure.

### *2.2.1 Reuse of HMI Code*

With regard to HMI, [19] emphasize its importance as a component of control software and its reuse; however, the proposed implementation of an exemplary development process does not include variability. Besides, [20] introduce a method for automatically generating HMIs from a process industry model, i.e., piping and instrumentation diagrams, allowing reuse through abstraction.

### *2.2.2 Reuse of Operating Modes in Control Code*

The technical standard OMAC PackML [21], which is predominantly applied in the field of packaging machines, complies with the levels of the physical model of ISA-88 and enables a condition-based realization of operating modes to connect machines from different manufacturers. The standard defines a set of machine states, which can be combined by transitions and arranged to state machines representing operation modes for controlling the desired machine behavior. Operation modes can be switched via a machine mode manager usually located at the highest level of the software architecture, which specifies in which states and by which transitions the change is triggered. In case of an error, different states are available to stop the machine, e.g., *aborting* for an immediate standstill or *stopping* for a controlled shut-down. Generally, the OMAC pattern supports the flexible control of CPPS as it enables the separation of the customer- or operation mode-specific process logic from the pure hardware control.

Based on PackML, [22] presents a modular software concept in which operating modes are implemented as automata. Thereby, the control code of a machine responds to a change of state, but HMI software is not targeted. Ladiges et al. [23] present the DIMA concept also using PackML for handling operating modes in modular process plants and include HMI functionality. Concepts for alarming and diagnosis are not yet included. With a focus on reengineering, [24] follow the approach of automatically converting control software into UML class and state diagrams. For documentation, the formalization also enables investigation of the software availability and performance. However, no special focus is laid on extra-functional software.

### *2.2.3 Reuse of Diagnosis and Error Handling*

For error handling, [25] and [26] work on a framework to investigate reliability in early development phases. Papkonstantinou et al. use feature models for functional aspects and integrate fail-safe (extra-functional) aspects to find valid combinations of configuration options [26]. Both integrate functional and extra-functional aspects at a high conceptual level in the process industry but do not consider concrete control engineering aspects. Güttel et al. [17] point out that there is still a gap to bridge regarding the automatic generation of extra-functional control software, including, e.g., communicating alarms to the user or diagnosis functionalities, and covering troubleshooting and displaying diagnosis results to the user.

Detailed code analyses of industrial PLC software examine the significance and strategies for implementing extra-functional control software and identified five hierarchy levels, which correspond to the ISA 88 levels [27]. Recent analyses show so-called design patterns in control software and experts from industry confirm their application for specific, extra-functional aspects [5]. Vogel-Heuser et al. analyzed common error handling strategies in industrial PLC code and developed performance metrics to evaluate real-time error detection and error coverage for the CPPS domain [28]. Besides, [29] developed a concept for reconfiguration in case of a PLC error and distinguished different methods of error reaction depending on error severity and system state. Similarly, [30] use a model-based approach to define restart points after a control error.

In classical control code, approaches with a designated error variable are used, whereby the error variable is checked and updated in every POU controlling an actuator of the machine, e.g., drives or cylinders (cf. variable *bStop* in Fig. 3). Errors are looped through the entire code: if a POU on the lowest level (e.g., *FB_Drive* for controlling a drive of the machine) sets its error variable, the error is reported to the higher levels (*PRG* or *OB1*).

In summary, various approaches target extra-functional control software and its reuse. However, up to now, no approach has focused on a comprehensive variability analysis in both, extra- and functional PLC software, to support systematic reuse.

### 2.3 Object-oriented IEC 61131-3 Concepts for Planned Reuse and Questionnaire-based Results on its Application in Industry

For enhanced reuse of control software, the OO extension

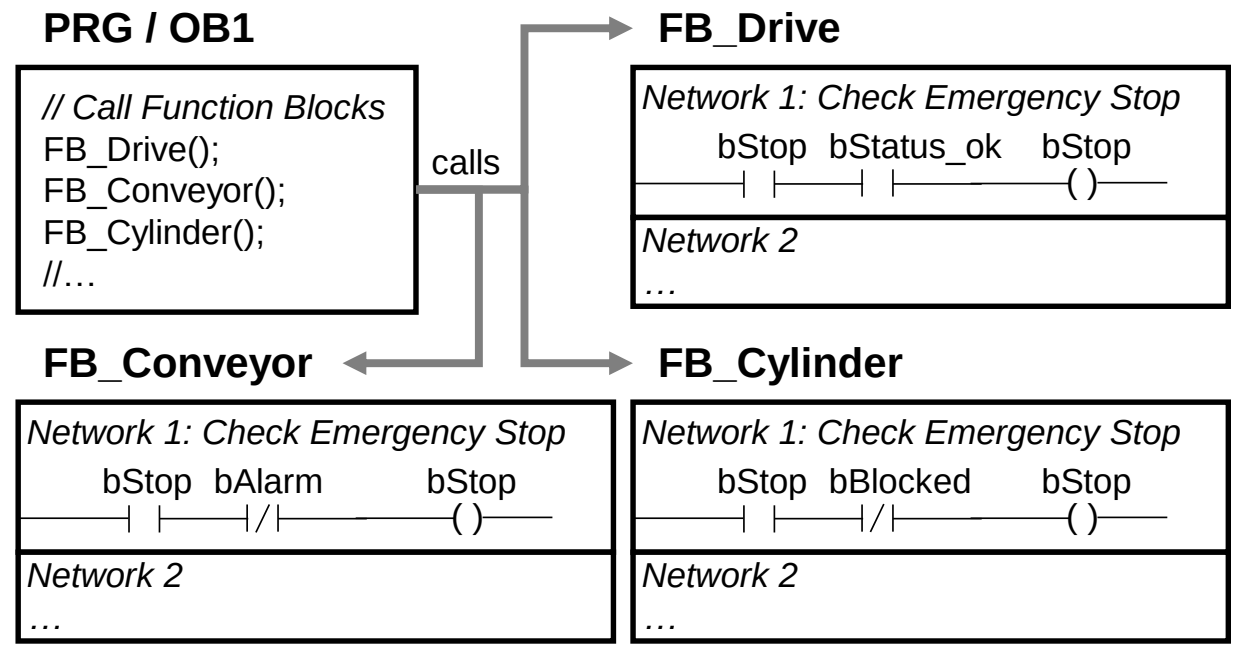

Fig. 3- Error handling (emergency stop) with a global variable accessed in all POUs responsible for actuator control (FBs programmed in ladder diagram) [6]

of IEC 61131-3 (OO IEC) introduces three new language elements, methods, inheritance and interface abstraction [8]. Thereby, methods enable the separation of an FB's tasks, e.g., initialization or error handling, in sub-routines. Further, inheritance supports the reuse of common code parts to ease building new variants with the keyword EXTENDS. Interfaces serve to improve the cooperation of developers in an IEC 61131-3 project by defining templates, which can be adopted and filled by FBs with the keyword IMPLEMENTS. While an FB can only inherit from one other FB, it may implement various interfaces [8]. Some PLC development environments also support the use of properties, which specify a defined access to a value and possess a pair of methods for reading (Get method) and writing (Set method) the respective value. This enables verifying data consistency and value adjustments. According to [31], using OO enables "a more efficient code reuse and increased safety and stability of software". To summarize, OO IEC provides a variety of rewarding solutions to improve reuse and quality of CPPS control software [6].

Although many platform providers meanwhile support the OO IEC concepts, a conservative, procedural programming approach is still predominant in most companies, which was confirmed by a recent survey: only 10% of the participating companies use OO IEC by default, 48% apply it partially and 42% do not use it at all [18]. Another recent questionnaire study confirmed this low usage rate of OO IEC with around 40% for machine and plant manufacturers [6,7]. Out of 61 participants, only 24 companies indicate to use interfaces/properties and OO IEC. It was further analyzed which extra-functional aspects their software modules include as standard. Ten out of ten companies include error handling and diagnosis as standard, nine operating modes, seven HMI and six include all three. A deeper analysis of category combinations like PLC type, interfaces, functionalities included within modules, checklists, version tracking and reuse strategy with its sub-categories universal module and libraries showed that participants answered unexpectedly: usage of interfaces/properties (without the limitation to IEC OO) reduces the usage of data exchange via global variables significantly. However, use of interfaces/properties would require the usage of OO IEC as a prerequisite. Companies answering the questionnaire neglected this prerequisite and must hence refer to other mechanisms. In summary, we can assume that the OO concepts are not widely used yet.

Overall, the high variability in CPPS and its efficient management still pose a challenge for the reuse of functional software. Further, extra-functional software and its strong dependencies on functional software make planned reuse of control software even more difficult. Although means for variant management and reuse exist, e.g., SPLE & OO IEC, they are not yet common in industry. To support planned reuse of variant-rich, (extra-)functional software, this paper presents two concepts in Sec. 4.

## 3 Background on Error Handling and Use Case highlighting Challenges in extra-

## FUNCTIONAL TASKS

First, an overview of typical steps and corresponding software design decisions in error handling is given (Sec. 3.1). Second, challenges of reusing extra-functional tasks for actuators are outlined using a lab-sized demonstrator.

### 3.1 Steps in Error Handling

Experience from more than ten industrial case studies and three in-depth interviews in companies of different industrial sectors confirms that the implementation of error handling as an extra-functional task is one of the key challenges in the design of control software architecture. More precisely, designing well-defined module or POU interfaces between the error handling software parts across modules and the functional software is a major challenge.

In an interview study with three companies from the packaging machinery sector, we investigated how the implementation of (extra-)functional tasks differs. For this purpose, we conducted separate, guided interviews with PLC software experts from each company, each lasting three hours. In these guided interviews, questions on the companies' design decisions regarding selected (extra-)functional tasks were targeted. After summarizing the findings, we verified the accuracy of the results in separate follow-up meetings with the experts from the individual companies. Regarding the implementation of error handling, commonalities could be observed despite different modularization strategies and boundary conditions. Based on those interviews, supplemented with the analysis of industrial code examples and large-scale questionnaire studies, four steps addressing different aspects in error handling have been identified that are necessary for most control software projects (cf. Fig. 4).

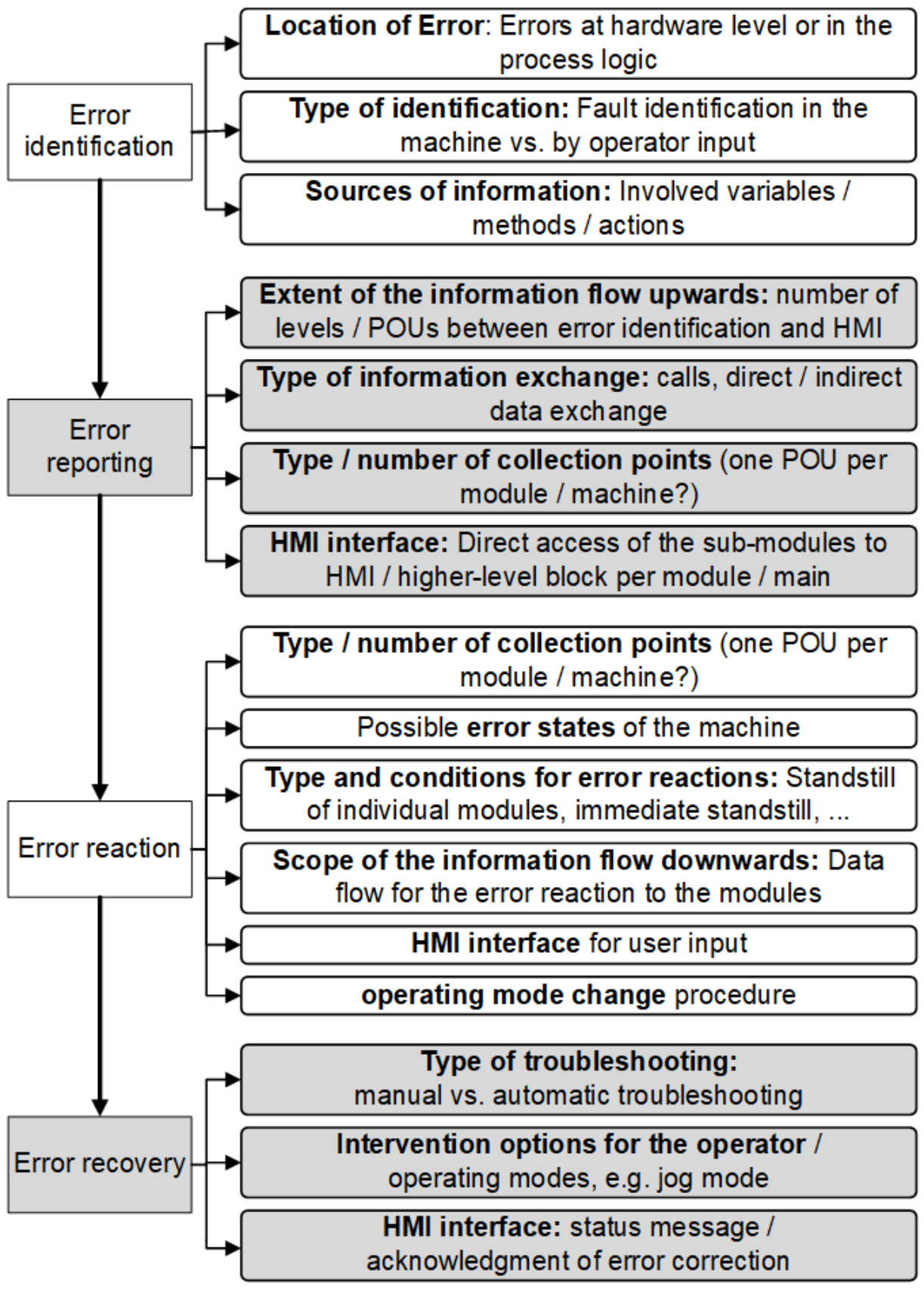

Fig. 4. Error handling steps and corresponding design decisions in control software

The first step is *error identification*, in which, e.g., a specific part of the software or the operator himself identifies a deviation from the intended behavior. It needs to be considered at which structural levels an error can occur (e.g., hardware misbehavior on actuator level or deviations in the process on higher levels) and whether these differences affect the implementation of the error identification.

Next, the error is communicated to other software parts during the *error reporting*. While there are often similarities in the implementation of *error identification*, *error reporting* shows a significant variability across companies, sometimes even across different machine types of the same company. Design decisions in this phase comprise, e.g., the type and amount of data exchange from the software module reporting an error to the software instance communicating the error information to the operator. Depending on the boundary condition or error type, the error can either be reported directly to the HMI, or collection points are implemented that can, e.g., trigger and pass on group alarms.

After reporting the error, the *error reaction* determines how the entire system or parts of the machine should react to a particular error. The reaction usually depends on the severity of an error. Categories differ between companies and machine types. A distinction between less critical (message, warning) and more critical categories (malfunction, error) can be found at almost every machine manufacturing company. Some companies even distinguish up to 70 error types leading to different machine reactions.

The last step of the error handling, i.e., the *error recovery*, covers the type of recovery process (e.g., automatically by the machine itself or via manual intervention by the operator) and the possibilities for the operator to interact with the system, e.g., via specific operation modes, such as a jog mode, or via the HMI.

### 3.2 Challenges in Documenting the Dependencies of variable, (extra-)functional Software Parts Utilizing a Lab-sized Demonstrator

In the following, a lab-size pick-and-place unit (xPPU) is introduced, which serves as a demonstrator for different ways of implementing (extra-)functional control software parts. The xPPU's main functionality is to sort work pieces (WPs) of different colors and materials. Using different types of sensors, the xPPU distinguishes the WP types and performs different intralogistic operations, such as separating, transporting or sorting WPs, with its actuators.

The mechatronic layout of the xPPU comprises different modules, each fulfilling a specific functionality of the WP handling process. Initially, the WPs are stored at the *stack*. Depending on the material, the *crane* transports the WPs to different end positions: metallic WPs require an additional processing step and are moved to the *stamp*, plastic WPs are placed directly on the *sorting unit*, which consists of a conveyor belt with several separators that push the WPs into different ramps depending on their type.

The xPPU's software is written in IEC 61131-3 languages and the architecture complies with the levels of ISA-88: the highest level, i.e., the xPPU module, refers to the *Unit,* while the modules, i.e., stack, crane, stamp, and sorting unit, are controlled by corresponding *Equipment Modules (EM).* Finally, *Control Modules* on the lowest level read sensor values and control actuators, e.g., cylinders. In the xPPU, cylinders are used in several EMs, e.g., at the stack to push out the WPs, at the crane to lift WPs, or at the conveyor belt to sort the WPs into ramps.

### 3.2.1 Extra-functional Tasks in Cylinders as Simple Variant-rich Standard Components

Cylinders represent a simple type of actuator in CPPS, which frequently occur in different module types. However, in industrial CPPS, even the control of such a simple component can get very complex and extensive. The following comparison of two FBs for cylinder control, one provided by a machine manufacturing company and one from the academic xPPU demonstrator, illustrates this: while the industrial cylinder FB provides a connection to the HMI, multiple error messages and different operating and testing modes, the academic implementation excludes these extra-functional aspects. The increased functionality of the industrial FB is reflected by the scope of its implementation, which is written in the language Ladder Diagram (cf. Fig. 3) and comprises 33 networks. On the other hand, the academic implementation comprises an FB with two actions in Sequential Function Chart with only three steps each. A comparison of defined variables illustrates the difference even clearer: while the academic FB uses only six variables, the industrial FB controls 150 variables, partly organized into User-Defined Types and Structures.

In both the academic and the industrial software, different types of cylinders occur, e.g., mono- and bistable cylinders (cf. Fig. 2), leading to variability in functional software and in the related HMI software, which directly accesses variables from the functional software part to illustrate the CPPS's status to the operator or maintenance staff. For variability management as a prerequisite to reuse, we propose to combine the family model of functional control software with the HMI family model via extra-functional control software, e.g., operating modes or error handling and diagnosis, being orthogonal to both (cf. Fig. 5).

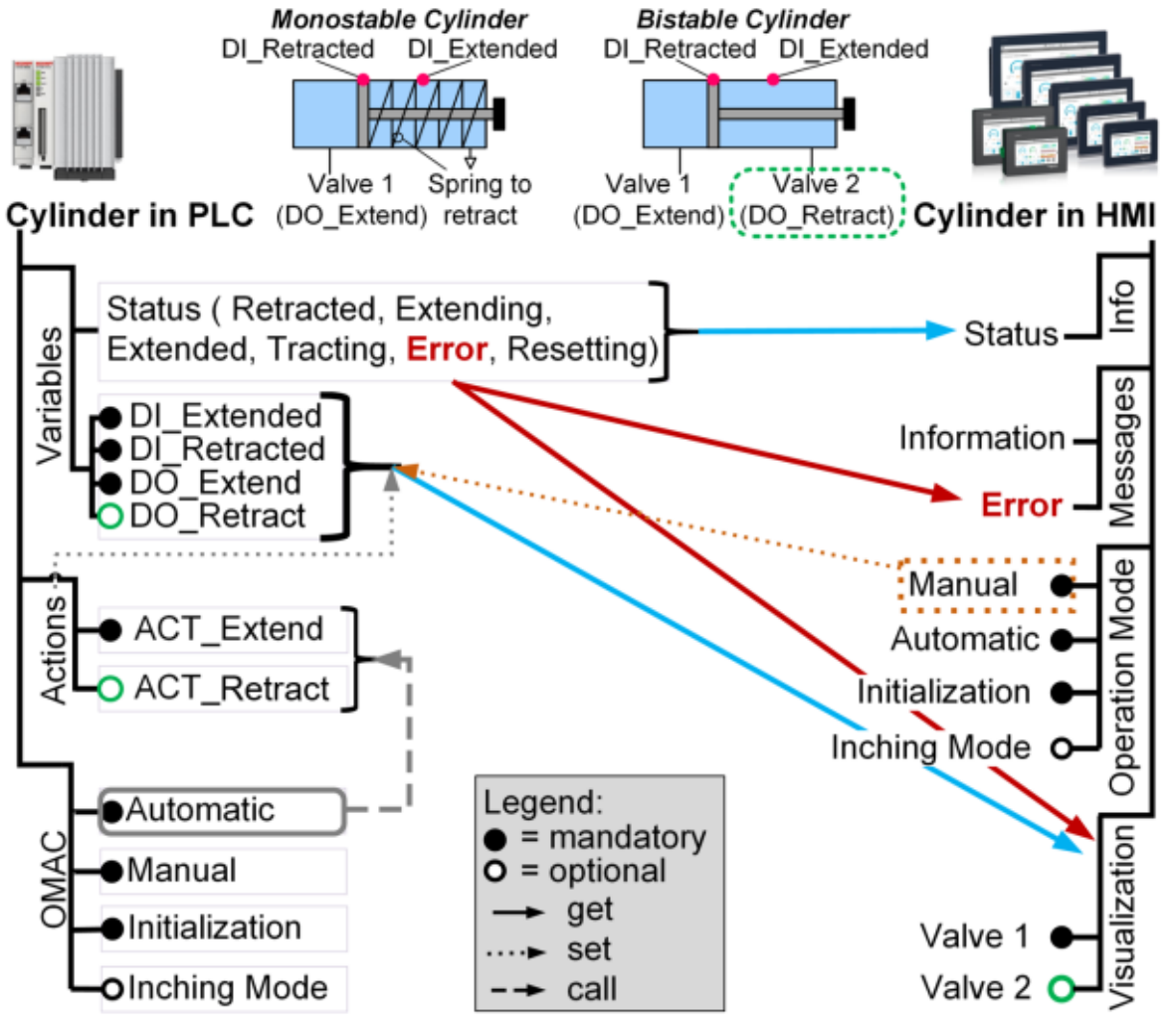

Fig. 5. Combining functional PLC code and HMI code via extra-functional tasks (operating modes, diagnosis and error handling)

A PLC family model (Fig. 5, left part), containing mandatory and optional parts to represent the mono- and the bistable cylinder, depicts variables, actions and OMAC actions (modes of operation). Thereby, OMAC actions use the cylinder actions for control, which in turn set the actuators (digital outputs *DO_Extend* and *DO_Retract*) and read the sensor signals. The HMI visualizes the cylinder (either with one or two valves) and its position (requiring recent sensor values) and displays general information about the cylinder status for the operator. Thus, it receives data from the PLC (variables *Status*, *DI_Extended*, *DI_Retracted*). In case of an error, the HMI control panel displays an error message, which is also read from the *Status* variable. In contrast, if the operating mode is set to manual (either by the PLC in case of an error or via HMI), the actuator variables are set from the HMI (mandatory *DO_Extend* and optionally *DO_Retract*, depending on the cylinder type). This example illustrates the close interdependencies between PLC and HMI for extra-functional aspects.

### 3.2.2 Extra-functional Tasks in Servo Actuators as Complex Variant-rich Standard Components

The concept for a more complex component, i.e., a servo actuator, will be introduced in the following. It exhibits more variations and is consequently more difficult but still aims at hiding variability behind generalized POU interfaces. It needs to be decided which variations need to be exhibited to the functional code, which can be hidden, and where: on the sub-component (servo drive) or component level (crane). Typical variations in servo actuators are, e.g.:

- Rotary vs. linear movement of the mechanics
- Rotary vs. linear design of motor
- Limited vs. unlimited movement range
- Feedback via potentiometer, incremental encoders or absolute encoders

Variability to be handled at the level of a servo drive is, e.g., whether an incremental or an absolute encoder is used to detect the position of a motor or mechanic or even a potentiometer as used for the crane (cf. Fig. 6). For the (extra-)functional PLC code, these details can be hidden, e.g., by representing a servo actuator as an *axis* that only has a reference position and an actual position.

Limited or unlimited movement range, in turn, is an example of variability to be dealt with on higher levels. Vendors typically provide FBs for controlling axes, e.g., in the form of library POUs, such as the FB *AxisModule* in Fig. 6. The *AxisModule* controlling the crane base can be configured according to the possible movement range of the crane (e.g., CraneBase.config.PositiveLimit := 360, CraneBase.config.NegativeLimit := 0). Through configuration, the *AxisModule* solves the problem of variability in the movement range and avoids sharing the compressed airlines of the vertical crane axis. Further, the HMI can read the *AxisModule* configuration (rotary/linear limited/unlimited movement) to automatically select the corresponding icon to visualize the current axis position (cf. Fig. 6).

Further, the variability of available operation modes needs to be considered: for the belt, e.g., an operation mode for endless moving would be suitable until the transported WPs reach the end position. For the crane, on the other hand, a positioning mode is required to move to specified positions according to user input.

Regarding error handling, the *AxisModule* provided by the vendor can identify standard errors, such as jamming of the motor or drag errors, i.e., a deviation between the reference and the actual position. In addition, other errors can be defined in the user-specific code depending on the type of controlled mechanics. For the belt, e.g., a module-specific error could be set if a WP is expected but not registered on the belt because, e.g., it has fallen off. This can be displayed to the user as a warning in the HMI. A more critical module-specific error could occur in the crane in case the product sensor of the gripper is not triggered after the WP has been gripped, e.g., because the sensor is defective or no product has been caught. This would result not only in a warning in the HMI but also in the standstill of the plant. In both cases, PLC and HMI need to consider the module-specific variability of potential errors.

# 4 Procedural and OO Concept to Ease Reuse of Error Handling

This section proposes two concepts representing implementation patterns for error handling in the software architecture, i.e., a procedural concept (Sec. 4.1) and a concept utilizing OO IEC (Sec 4.2). The section closes with a comparison of both concepts in Sec. 4.3.

## 4.1 Procedural Concept for Error Handling

Software architecture and design play a key role in reusing control software, especially considering variant-rich and extra-functional parts. The following two subsections introduce a procedural concept for reusing (extra-)functional control software and a concept tailored to error handling.

### *4.1.1. Procedural Concept to Enable Reuse of Functional and Extra-functional Control Software*

To support the development of high-quality software for this programming style, some platform suppliers provide templates that their customers can use as patterns for modularizing their control software. Next, the template of a renowned platform supplier is introduced as an example for such a procedural concept, which is primarily designed for packaging machines and is compliant to OMAC and to the Weihenstephan guidelines for machine standardization.

The template represents a universal control software project, which can be copied and adapted by the users to integrate customer-specific software components. Basic extra-functional tasks, such as OMAC-compliant operation modes, diagnostic mechanisms and error handling, are already included and can be configured for user-specific applications. The template follows a modularization approach oriented towards the functional structure of the machine, which is usually also reflected by the physical machine layout.

Functional units are controlled by *Equipment Modules (EMs)* comparable to the EMs defined by ISA-88 and represent reusable software units to enable a scalable, modular project architecture. An EM comprises one or more *Application Function Blocks* (AFB), which refer to common automation tasks and machine functionalities and are usually parameterizable for different machine or task variants. Beyond the pure functionality defined within AFBs, EMs include a standardized module interface as well as extra-functional tasks on module level, such as switching of operating modes, error handling, and diagnostic functions, which makes them ideal basic components to create a well-defined software architecture. By structuring the software project into EMs and providing flexible mechanisms to link them via standardized module interfaces, the template enables users to encapsulate and reuse entire machine functions. Users can individually integrate new EMs, e.g., by copying and modifying the provided EMs from the template, encapsulating and using provided EMs as aggregation modules or writing new EMs from scratch and linking them to the provided template infrastructure.

Commonly used software functionalities and extra-functional tasks are provided as reusable AFBs or EMs in the form of application-specific libraries containing most of the functionalities required for different aspects of packaging, e.g., forming, filling, cartoning, labeling, or pick-and-place applications. Additionally, the libraries provide mechatronic functions typically used in production machinery and in handling, assembly, and sorting systems. The library POUs have been tried and tested in practical use, which also eases certifying machines and software.

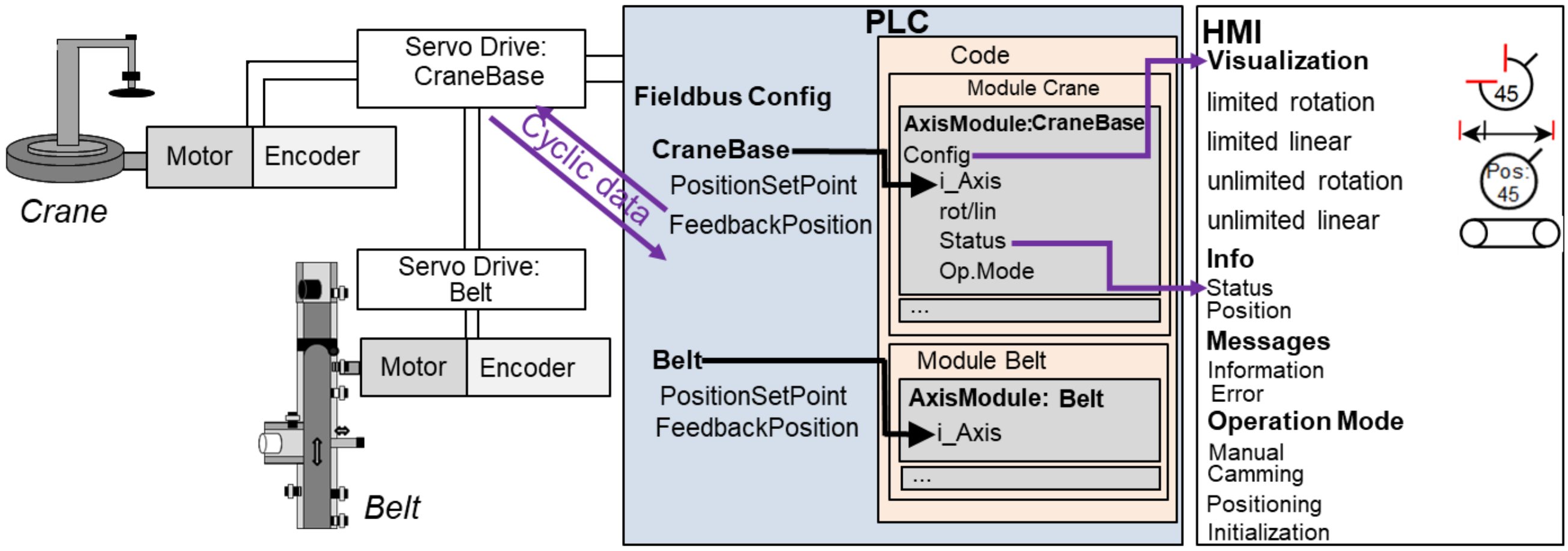

Fig. 6. Connection of the motor of the mechanical belt, servo drive, and functional PLC software, up to HMI

Additionally, the template architecture includes communication with the HMI, machine level command processing, operation mode management, as well as a machine-wide error handling and logging. In summary, the template provides a basic module structure that enables the reuse of functionality within a project and across projects or machines. In addition, clearly defined module interfaces between functional and extra-functional software enable to flexibly adapt and extend the modules for different types of application projects. The function-oriented structure supports the development of transparent and easy-to-use programs.

Although the template is primarily designed for a procedural programming style, the programming environment supports OO-IEC and some customers extend the provided library modules using inheritance.

### *4.1.2. Procedural Concept to Implement Error Handling*

The following section outlines the required characteristics and the design of the error handling concept in the procedural template (cf. Fig. 7) introduced in Sec. 4.1.1, which provides the following functions:

- *Error Identification:* Identification is possible on any architectural level (in this case ISA-88) by setting a Boolean variable.
- *Error reporting:* Error information is forwarded and summarized for the HMI using a method that reads this variable and writes the error information (e.g., error message and number) to a central location, e.g., a global error list connected to the HMI.

Communication of the error reaction from the HMI to the lowest levels of the software architecture is the most comprehensive part of the concept and includes the following aspects:

- *(1) Possible reactions:* Standard reactions for all axes are provided (ranging from an immediate asynchronous process stop to a controlled stop at the end of a process cycle). Additionally, application-specific reactions can be added.
- *(2) Broadcast to sub-modules:* According to the operator's chosen reaction (HMI), a central POU on the highest level of the software architecture reads all stored errors in the error list and broadcasts the reaction to all underlying EMs, which then spread the reaction to their axis modules.
- *(3) Reaction specification for individual sub-modules*: In case, e.g., modules must continue to run even in certain error states, this is achieved via so-called reaction matrices (Fig. 7). These matrices can contain more entries than potentially possible error reactions in the machine, which means that not all entries cause a reaction. This can be used to trigger reactions specifically for individual modules

Finally, the concept also defines the operation mode switch to resolve the error:

- *Error recovery:* If an error reaction is triggered, a corresponding bit is set at machine level, e.g., to terminate the automatic mode and to switch to another operating mode (e.g., jog mode).

This template-based, procedural concept is used by two of the packaging machine manufacturing companies, which took part in the interview study introduced in Sec. 3.1. Both companies implement their error handling using the template provided by the platform supplier and enlarge it according to their needs. Thereby, the companies need to consider specific boundary conditions, e.g., requirements for sterility (cf. [6] for additional details).

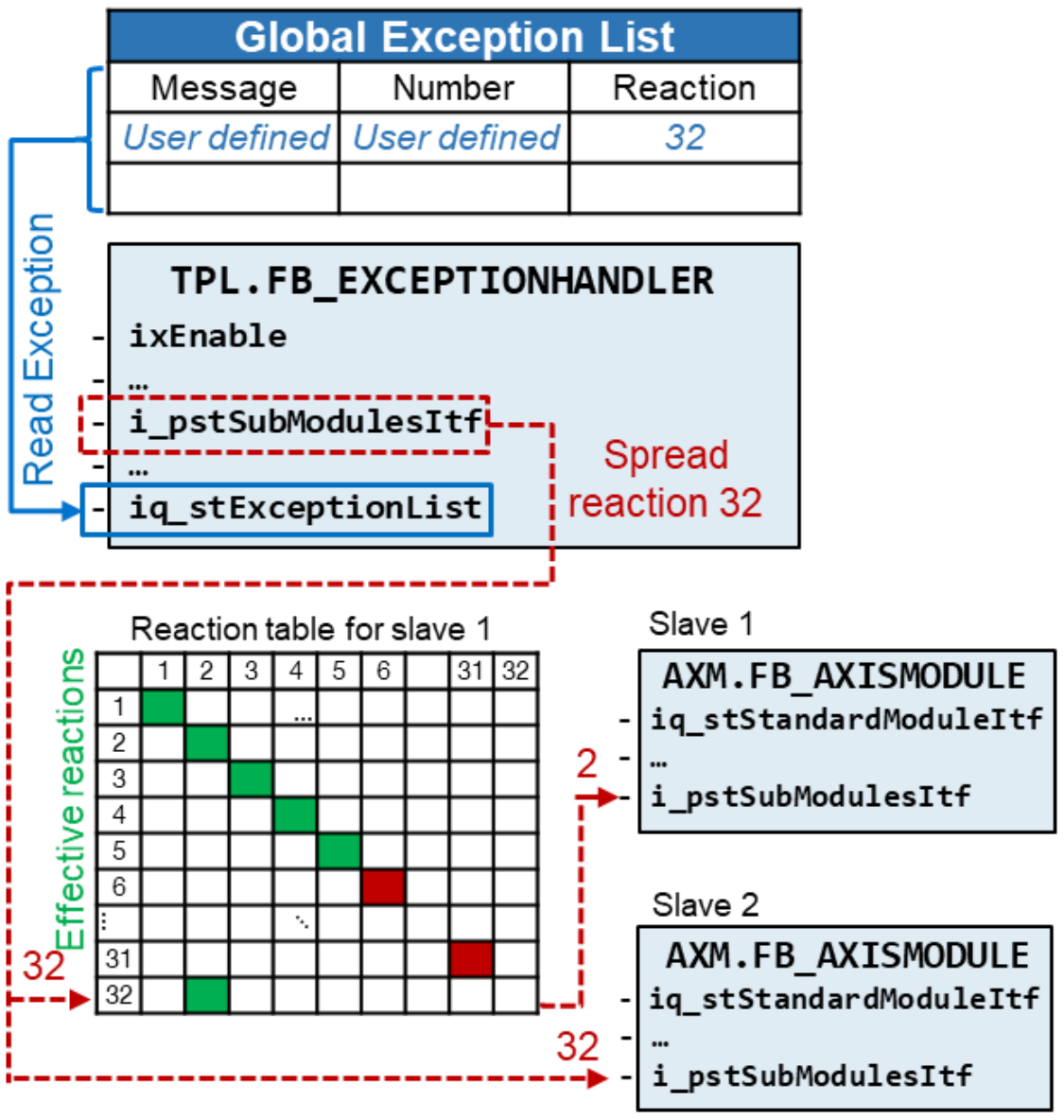

Fig. 7. Reaction matrix to broadcast a reaction (e.g., reaction 2) only to a specific sub-module by writing the reaction to one of the in-effective lines for the module (e.g. 32 for slave 1, which only reacts to 1-5).

## 4.2 Using OO IEC 61131-3 Concepts

Despite challenges and obstacles to use OO in machine and plant manufacturing companies [18], the concept is beneficial for variant-rich and extra-functional software tasks, as introduced below in Sec. 4.2.1. A callback facilitating the standardization of error handling is presented in Sec. 4.4.2.

### *4.2.1 OO IEC to Ease Reuse of Variant-rich or Extra-functional Control Software*

As highlighted by [6] and [8], the usage of OO IEC should tremendously ease modularity, reuse and managing variants and versions. The OO IEC concepts provide the means to design a well-defined software architecture, which supports the reuse of common parts, prevents code clones and enables standardized interfaces. At the example of the cylinder variants from Sec. 2.1, we illustrate the use of OO IEC and its benefits (cf. Fig. 8). Additional theoretical background on the OO IEC concepts can be found in [6].

Concerning variability within the software, the cylinder family model (cf. Fig.2) illustrates common and variable parts. The OO IEC concept *Inheritance* supports the reuse of common parts used by several modules (e.g., *METH_Extract* used by both cylinder variants). They are defined once in the base class and then inherited by the specific module variants (cf. Fig. 8, *Method A* defined in the base and inherited by *FB_BistableCylinder*, which defines an additional *Method B* for retracting). This reuse reduces the programming effort and facilitates module maintenance, e.g., in case of errors, when a correction of the inherited parts is necessary and can be performed once in the base class.

For information hiding (to transfer or query values to/from an FB from outside), the OO IEC concept *Properties* enables the access of defined variables. If values that are available in each module need to be obtained, e.g., the current state of the cylinders, these values can be accessed externally using properties. To avoid creating properties required by various modules multiple times, they are defined in a base module (cf. Fig. 8, *FB_Base*, Property *State*). The FBs of different modules can then use these properties via inheritance, e.g., in *FB_BistableCylinder*.

In some cases, inheritance is insufficient as an FB can only inherit from one Base-FB but not from multiple bases. OO IEC *Interfaces* address this challenge: an OO interface has only declaration parts, in which methods and properties are defined. If an FB implements an interface, it must adopt all methods and properties contained therein (cf. Fig. 8, *FB_BistableCylinder* inherits the Methods from Interface *IUnit*) and implements them. For switching between different operating modes in various modules of a CPPS, interfaces can be used. The operating modes are defined as methods in an interface and then individually implemented for all module FBs. With a generated module list, the methods of all modules are called in a for-loop, enabling an efficient operating mode change.

### 4.2.2 OO Concept with Callback Pattern to Communicate an Error and Its Severity

The idea behind a callback or "function-as-parameter" [32] is to hand over an executable software part as a parameter to another software part, which is expected to execute (call back) the passed software at a given time. An application example for error handling is introduced to demonstrate an implementation of a callback in OO IEC.

As described in Sec. 4.1.2, standard error handling and application should be separated. A standardized library FB for error handling cannot call a POU of the application-specific software part or access global variables. For this purpose, the concept of an *abstract OO interface* can be used to interact with a standardized software part. Below, details of this concept with callback pattern are illustrated.

The first part is an OO interface containing a method (cf. Fig. 9 top, interface *ITF_ExtFunc* with *Method_A*). An FB dedicated to addressing extra-functional tasks, in the example *FB_ExtraFunctional*, implements this interface and programs the functionality of *Method_A*, e.g., a standardized format for collecting and forwarding information regarding an error. Thereby, *Method_A* represents functionality that needs to be passed to all actuators to enable a standardized way of implementing error handling. For this purpose, in the application-specific part, the main program declares an instance of *FB_ExtraFunctional*, i.e., *FB1* in Fig. 9. As an example for the control of a standardized hardware component, *FB_BistableCylinder*, which controls bistable cylinders, is used. For passing the implementation of *Method_A* to *FB_BiCyl*, an instance of *FB_BistableCylinder*, the abstract interface is used. More precisely, the FB has a variable *var* and a property *Itf*, both of the type *ITF_ExtFunc.* In the implementation of the main program, via the Set()-method of property *Itf* the instance *FB1* of *FB_ExtraFunctional* is passed to the variable *var* of *FB_BiCyl*. Since *FB_ExtraFunctional* and, thus, its instance *FB1* implements *Method_A*, the Cylinder-FB instance can now access the implemented *Method_A* via its variable *var* (cf. implementation of *FB_BistableCylinder* calling *Method_A* via its variable *var* in Fig. 9). Thus, the handover of an interface's method (i.e., *Method_A* of *ITF_ExtFunc*), which is programmed within an FB implementing the interface (i.e., *FB_ExtraFunctional*), is enabled via a property (i.e., *Itf* of *FB_BistableCylinder*) initializing a variable of the type of the respective interface (i.e., *var* in *FB_BistableCylinder*) with an instance of the implementing FB (i.e., *FB1*). In short, this interface concept allows to hand over pointers to implemented methods via a parameter, i.e., a callback. Thereby, the passed method is implemented only once and every callback triggers the execution of the single method imple-mentation (cf. Fig. 9 top right, storage area of the PLC containing the only available implementation of *Method_A*). Please note, due to simplicity reasons, the example contains only one interface with one method. Actually, an FB, e.g., *FB_ExtraFunctional*, can implement more than one interface with more than one method each.

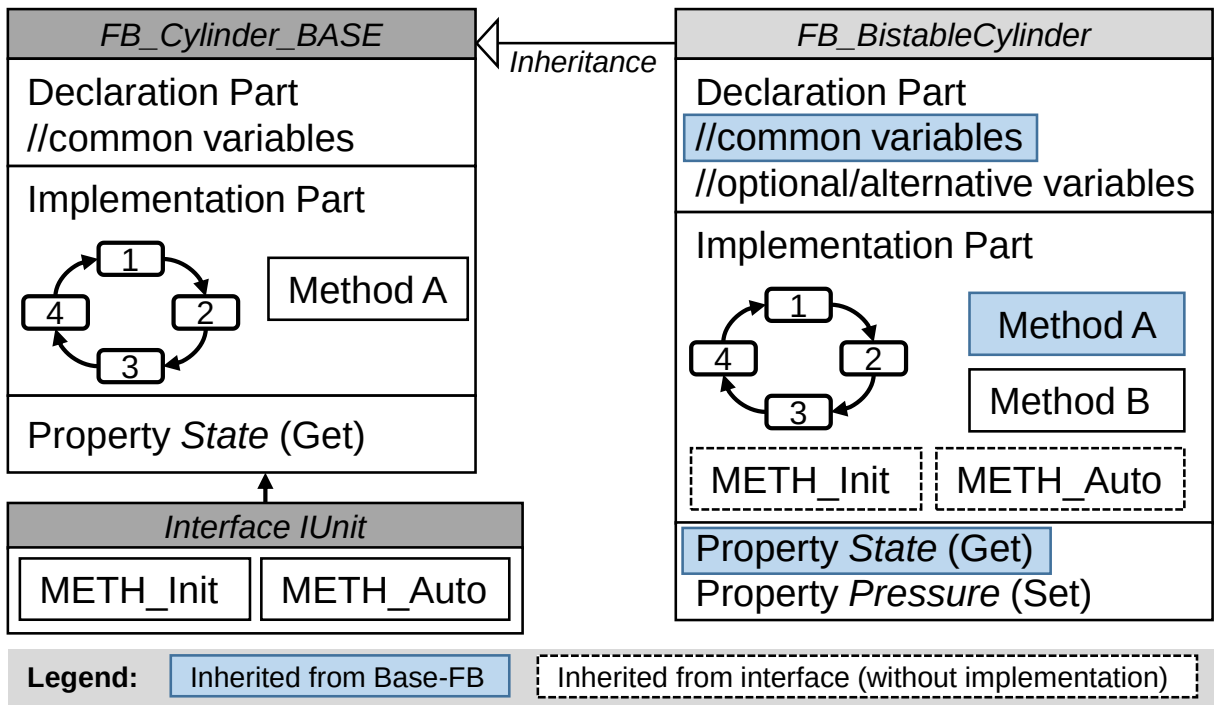

Fig. 8. Using OO IEC for eased reuse of (extra-)functional software via inheritance and interface [6]

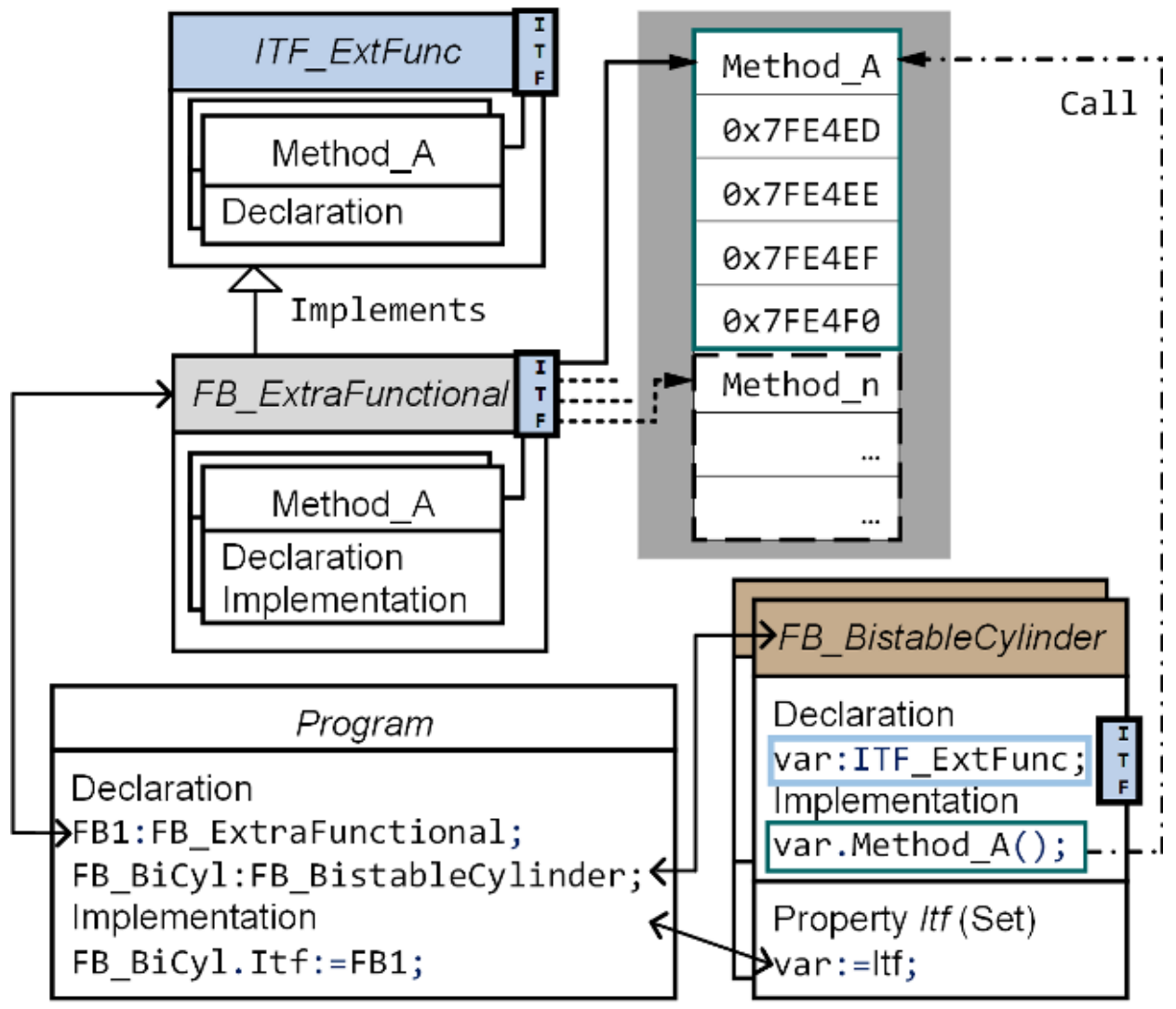

Fig. 9. Fault handling with the Callback pattern using OO IEC (brown: application FB; grey: Standard FB for extra-functional , blue: OO interface to provide standardized module interfaces)

With an exemplary use case of the described callback pattern, its benefits are illustrated: the passed method *Method_A* could be used to process analog values of an FB for hardware control from outside. In this scenario, the hardware module FB supplies the analog value, which is converted via external access, with the method implementing the desired conversion. In this way, not only can data be processed, but also functional behavior can be passed to an FB from the outside retrospectively, whereby the exact behavior may not yet exist when the FB for module control is developed or may have to be adapted for the desired, application-specific behavior. Thus, a behavior can be foreseen during the module FB implementation without knowing the exact characteristics of this behavior and the possibility of changing or adapting this behavior to evolving or application-specific boundary conditions.

According to this principle, an OO interface can be used as an error manager to collect various information from different modules, which can be treated as black boxes. When using an OO interface as an error manager, the error reporting is integrated into all actuator FBs, e.g., the cylinder FB. For this purpose, each cylinder FB instance gets an *abstract* interface of the error manager in the initialization step of the main program, similar to Fig. 9 (code line *FB_BiCyl.Itf:=FB1;*). This OO interface can be used to report errors. The diagnosis of the errors happens in the cyclic call of the cylinder FB. This procedure provides detailed error messages – there is no handling of errors outside the call – and it allows the additional output of detailed information, including severity. Furthermore, to ease reuse, independent of the callback pattern, the automatic sequence is separated from the organizational sequence. More precisely, the advantage is that the automatic sequence, i.e., the process logic, which is made anew for each plant, does not contain any error handling.

The callback pattern is commonly applied in high programming languages and has a great potential for easing the reuse of extra-functional control software parts. It is already used for programming CPPS in two different ways. First, the pattern is contained in libraries provided by platform suppliers. Thus, customers using these libraries use the callback pattern without noticing it. Second, although OO IEC is not yet widely spread in CPPS programming, according to a platform supplier, a handful of its customers have converted their control software completely to OO IEC and thereby use the callback pattern multiple times. These customers report that they highly appreciate the benefits of OO IEC.

### 4.3 Comparison of the Conventional and OO Concepts Regarding Their Suitability for Implementing Extra-functional Aspects

After introducing two error handling concepts, their advantages and disadvantages are evaluated (cf. Table I). Both are illustrated abstractly and mapped and related to the hierarchy levels of [27] in Fig. 10.

Generally, the procedural concept described in Sec. 4.1.2 is considered easy to understand with basic IEC 61131-3 programming knowledge. Further, when dealing with legacy software systems that do not yet (fully) support OO IEC, it is beneficial for introducing a well-structured and reusable error handling strategy, which includes, e.g., location and severity of an error, e.g., in the monostable cylinder (cf. Fig. 10). Further, the template allows enlarging standard module errors with customer- or application-specific errors in the *Central Exception List* and the *Reaction Matrix* (cf. Fig. 10).

Despite these advantages, the conventional concept holds some drawbacks, e.g., the dependability on a global structure (cf. *Central Exception List* in Fig. 10, which must be passed by the caller with each call of *FC_SetException* by accessing a global variable). In the case of running the control programs of independent machines on one PLC to save costs by reducing the automation hardware, the error handling or other extra-functional aspects of these machines cannot be kept separate due to the global structure (cf. matrix and central exception list in Fig. 10). Moreover, when using this template, the implementation of error handling and the actual module control, i.e., extra-functional and functional aspects, are mixed. Additionally, the framework provides other functionalities apart from error handling, resulting in a high amount of information to be exchanged between framework and modules. Due to the mixture of (extra-)functional aspects and the tight coupling between framework and modules, module reuse is strongly bound to the template and cannot be used easily in a different framework or for testing purposes, e.g., unit tests. Another disadvantage results from the lack of encap-sulation, which allows modules to access internal execution details, utility functions or temporary storage data of the framework. More precisely, a module, e.g., *Crane* in Fig. 10, has complete access to the contents of the global exception list. Only by convention should it neither read nor write the contents of the list, but instead should only call *FC_SetException* and pass the *Central Exception List* (cf. Fig. 10). Finally, there is no check if all necessary information in the reaction matrix is filled in completely or correctly, which may result in a field not being filled in or in an incorrect assignment of (only weakly type-tested) pointers. Despite numerous integrated protective measures, it is not possible to mitigate all these risks completely.

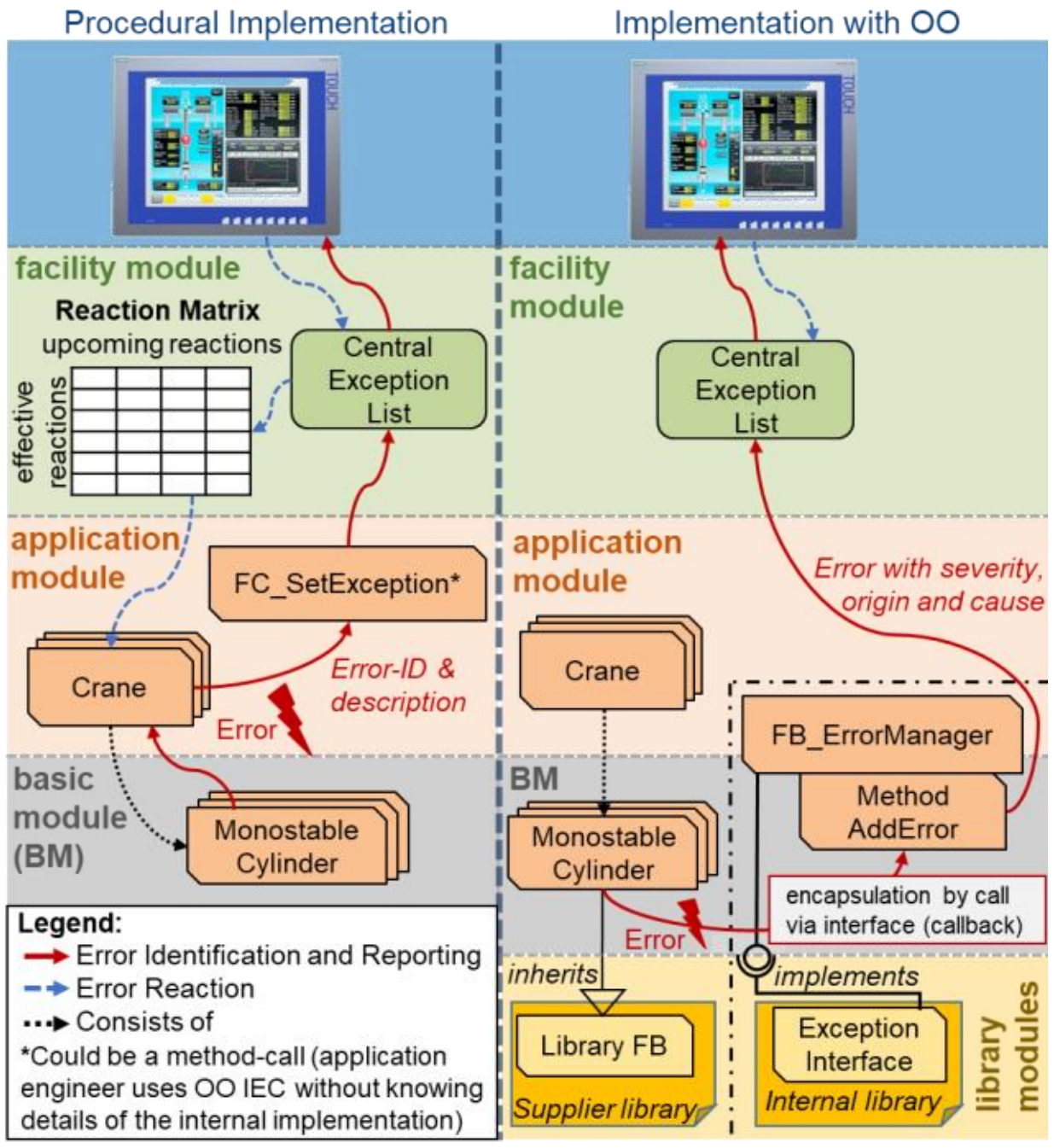

Fig. 10. Comparison of the error identification and reporting

In the OO concept, similar to the conventional concept, errors are also gathered on top-level in a central error list (cf. Fig. 10, right). If an error occurs, the affected module calls the method of the central error management FB (cf. *FB_ErrorManager*) to add the error to the list, including severity, origin and cause, for visualization at the HMI. Since the modules are updated every cycle and report the error independently, no cyclical query is necessary. Once an error is detected, the module's property *hasError* is set to true to report the error. Via a further interface, a module can query the status of the neighboring modules and, depending on their status, determine a suitable error reaction, e.g., immediate standstill or stop after the next cycle.

Regarding comprehensibility, the OO concept might initially be hard to understand for programmers with little or no background in OO programming principles. Due to the higher amount of required elements, i.e., the definition and declaration of an error handling FB with a method, an interface and a property (cf. Fig. 9 and 10), and their interconnection via a callback, the concept may initially appear more complex than the conventional concept. Further, to use the concept, the respective system needs to support the OO extensions of IEC 61131-3, which might not always be the case when dealing with legacy systems.

Like the conventional concept, the OO concept has a central method for adding an error, which enables a uniform structure of the error information, including severity and location. However, in contrast to the conventional concept, using *FB_ErrorManager* does not require accessing global variables as each module has access to the *Central Exception List* via the interface reference to *FB_ErrorManager* and its method (cf. Fig. 9). Especially regarding extra-functional aspects and their reuse, the OO concept shows several advantages: the OO interface links the extra-functional aspects, i.e., the error handing implementation (cf. *FB_ErrorManager* with Method *AddError* in Fig. 10), and the functional aspects, i.e., the module control (cf. Fig. 10, application module *Crane* and basic module *Monostable Cylinder*), while it allows to implement them separately. This clear separation of concerns enables to exchange the error handler if required. At the same time, the extra-functional implementation within the method is only used by the modules, e.g., the *Monostable Cylinder*, but cannot be modified. Further, the modules are usable within a different context since the OO concept, unlike the conventional approach, is independent of the template. Another benefit of the OO concept is that the central error manager (cf. *FB_ErrorManager* and its method in Fig. 10) can be extended by additional functionalities through inheritance. In this case, the current error manager serves as the base class for a new, extended one. This allows legacy modules and new modules to be used in combination, whereby the legacy modules only use a limited functionality of the new error manager. Moreover, CPPS- or application-specific errors can be included. Further, the usability of the OO concept is better than the conventional concept as it can be used less incorrectly. On the one hand, the implementation with interfaces and methods prevents forgetting to fill in the required information, e.g., data regarding severity, origin and cause of an error when calling *AddError*. On the other hand, the risks of wrong assignments are reduced.

Table I: Comparison of conventional and OO pattern regarding their suitability for implementing extra-functional aspects (+ advantageous, o: medium, -: disadvantegeous)

| | **Conventional Concept** | **OO Concept** |
|---|---|---|
| Comprehensibility (with classical IEC 61131-3 background) | **+** (basic knowledge sufficient) | **o** (basic knowledge not sufficient) |
| Legacy system support | **+** (supported, no OO IEC required) | **-** (OO IEC not supported by all systems) |
| Customer- / application-specific errors | **+** (template can be enlarged) | **+** (enlargement through inheritance) |
| Error severity & location | **+** (uniform, structured information) | **+** (uniform, structured information) |
| Reuse in different context | **-** (modules highly depend on framework) | **+** (template-independent) |
| Modularization | **-** (no clear separation of extra- and functional aspects; no separation of extra-functional aspects with several PRGs in one PLC) | **+** (interface serves as link between error handler (extra-functional) & module control; exchange of error handler possible) |
| Encapsulation of extra-functional aspects | **-** (implementation of extra-functional aspects can be manipulated from the module) | **+** (modules cannot manipulate extra-functional implementations) |
| Usability | **o** (risk for incomplete or incorrect use) | **+** (low risk for incomplete or incorrect use) |

In summary, the OO concept is both, template-independent and easy to use due to the low risk of errors caused by improper use. It enables a clear separation of concerns, which allows to enlarge or even exchange extra-functional implementations, e.g., error handling.

A first evaluation with 131 bachelor students from TUM's mechanical engineering faculty confirmed that the use of OO IEC is initially challenging. Students solved two small OO IEC programming tasks on paper in the exam: first, using an interface and methods. Of the 131, 123 tried to solve the task; of these, 60% gave wrong or incomplete answers for the interface use and 28% had difficulties in the programming of methods. The second task targeted inheritance from an FB, including a call of the inherited method and its extension. 115 students answered, of these 10% made mistakes regarding inheritance, 24% used the SUPER-call to execute the method wrong or left it out and, surprisingly, 37% had problems in defining methods. Generally, students found it challenging to distinguish between interfaces and inheritance. However, for a reliable evaluation in a future step, the stakeholders *module* and *application developers* will be asked to solve tasks targeting characteristics such as flexibility, evolvability and maintainability of the control software and the related CPPSs.

## 5 Conclusion and Future Work

This paper introduced the challenge of handling variability in CPPS, which is included in multiple, partially overlapping views, spanning across functional and extra-functional PLC software, which additionally has high interdependencies to HMI software. Up to now, research has focused on either functional or selected aspects of extra-functional code. Considering the variant management as a prerequisite to planned reuse of both, we introduced the challenge to connect family models of functional code and HMI code related to the orthogonal extra-functional code.

Additionally, we proposed two approaches for implementing the extra-functional task *error reporting* as a central task linking variable functional hardware control with other extra-functional aspects like operation mode change and interaction of operators via HMI. The comparison of the proposed procedural and OO-based concepts for *error reporting* illustrated the advantages of OO IEC when implementing extra-functional tasks. Further, OO simplifies the separation between functional and extra-functional code and enables different developers to program the two parts independently. This software architecture also supports the flexible adaption of extra-functional aspects and the flexible integration of different variants of functional control based on standardized interfaces. In future work, approaches for implementing other extra-functional tasks with OO IEC will be developed and analyzed.

Of course, the introduced two concepts are already used in the design of control software of machine and plant manufacturing companies, but not as widespread as possible and expected. Consequently, obstacles in usage would need to be revealed in future work using interviews on the one hand and training sessions on the other.

## Acknowledgment

This work was partially supported by two grants from the German Research Foundation (DFG) with project numbers 335427442 and 451550676. The corresponding author of this paper is Birgit Vogel-Heuser (vogel-heuser@tum.de).

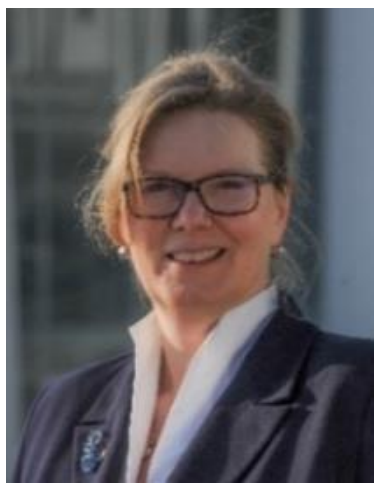

**Birgit Vogel-Heuser** (Senior Member, IEEE) received a Dr.-Ing. degree in Electrical Engineering and a Ph.D. degree in Mechanical Engineering from RWTH Aachen University, Aachen, Germany, in 1991. She was involved in industrial automation with the machine and plant manufacturing industry for nearly ten years. After holding different chairs of automation she has been Head of the Automation and Information Systems Institute at the Technical University of Munich, Munich, Germany, since 2009. Her current research focuses on systems and software engineering. She is member of the acatech (German National Academy of Science and Engineering), editor of IEEE T-ASE and member of the science board of MSRM at TUM.

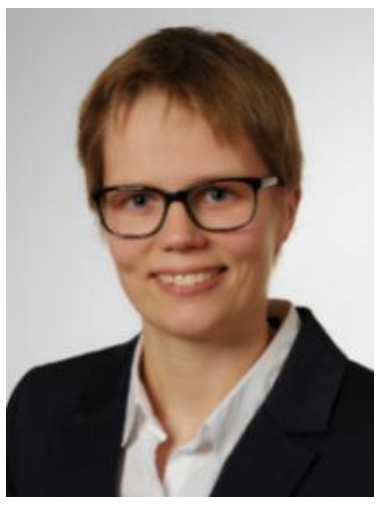

**Juliane Fischer** received an M.Sc. in Mechanical Engineering from the Technical University of Munich (TUM), Munich, Germany in 2017. She is currently pursuing a Ph.D. at the Institute of Automation and Information Systems at TUM. Her main research interests are the design of modular, reusable control software and methods from the field of static code analysis to enhance the reuse of variant-rich, legacy control software via identification of potentials for software improvement.

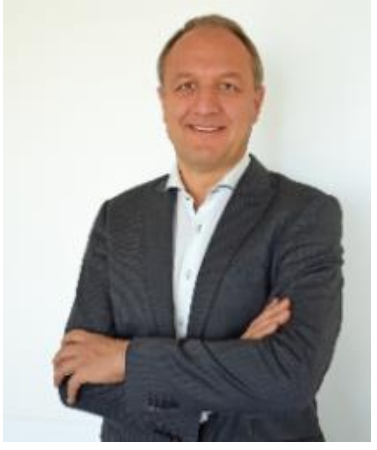

**Dieter Hess** holds a Diploma in Information Sciences from Technical University of Munich (TUM). He is co-founder and CEO of the CODESYS GmbH and responsible for product development, marketing and sales. Ever since the CODESYS GmbH included the object-oriented extensions of IEC 61131-3 in their development environment, he has been working on guidelines and support for its use to improve software quality, e.g., reusability, maintainability and testability. Further, CODESYS GmbH is involved in several research projects to ease engineering of control software.

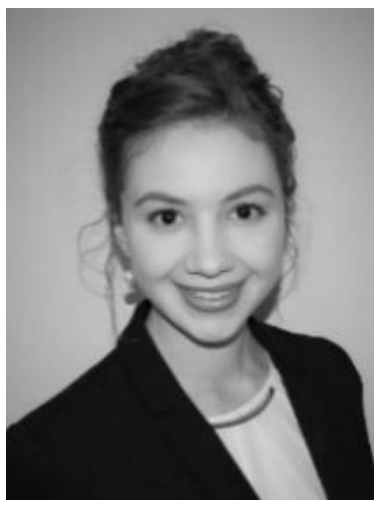

**Eva-Maria Neumann** received an M.Sc. in Mechanical Engineering from Technical University of Munich (TUM), Munich, Germany in 2018. She is currently pursuing a Ph.D. at the Institute of Automation and Information Systems at TUM. Her main research interests are static code analysis and metrics to quantify control software quality and the function-oriented design of modular control software architectures to enhance its reusability.

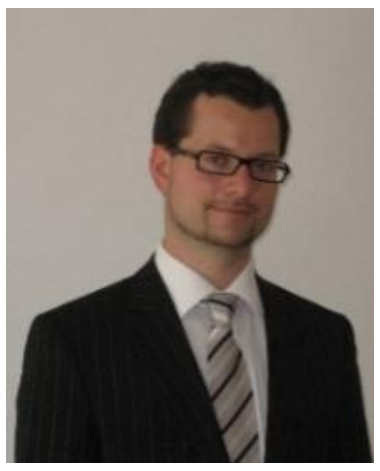

**Marcus Würr** holds a Diploma in Information Sciences from the University of Applied Sciences Würzburg-Schweinfurt. He currently works at Schneider Electric as an architect for embedded motion control solutions. His focus includes design of motion functions and modules for the packaging domain. He has been working on utilizing and improving the object-oriented extensions to 61131 frameworks. This has resulted in research contributions and pioneering functions for the EcoStruxure Machine Expert and Machine Advisor platforms.